\documentclass[prl,twocolumn,showpacs,groupedaddress]{revtex4}
\usepackage[dvips]{epsfig}
\usepackage{subfigure}
\usepackage{float}
\usepackage{amsmath}
\usepackage{amssymb}
\usepackage{amsfonts}
\usepackage{rotating}
\usepackage{euscript}
\usepackage{subfigure}
\usepackage{enumerate}
\usepackage{hhline}
\usepackage{threeparttable}
\usepackage{supertabular}

\begin{document}

\bibliographystyle{apsrev}

\title{\bf{Experimental demonstration of a high quality factor photonic crystal microcavity}}

\author{Kartik Srinivasan}
\author{Paul E. Barclay}
\author{Oskar Painter}
\affiliation{Department of Applied Physics, California Institute of Technology, Pasadena, CA 91125, USA.}
\email{phone: (626) 395-6269, fax: (626) 795-7258, e-mail: kartik@caltech.edu}
\author{Jianxin Chen}
\author{Alfred Y. Cho}
\author{Claire Gmachl}
\affiliation{Bell Laboratories, Lucent Technology, 600 Mountain Avenue, Murray Hill, New Jersey 07974, USA.}

\date{\today}

\begin{abstract} Sub-threshold measurements of a photonic crystal (PC) microcavity laser operating at $1.3$ $\mu$m show a linewidth of $0.10$ nm, corresponding to a quality factor ($Q$) $\sim1.3\text{x}10^{4}$.  The PC microcavity mode is a donor-type mode in a graded square lattice of air holes, with a theoretical $Q \sim 10^{5}$ and mode volume $V_{\text{eff}} \sim 0.25$ cubic half-wavelengths in air.  Devices are fabricated in an InAsP/InGaAsP multi-quantum well membrane and are optically pumped at 830 nm.  External peak pump power laser thresholds as low as $100$ $\mu$W are also observed.        

\end{abstract}

\maketitle

Optical microcavities are typically characterized by two key quantities, the quality factor ($Q$), a measure of the photon lifetime for the relevant optical cavity mode, and the modal volume ($V_{\text{eff}}$), a measure of the spatial extent and energy density of the mode.  Photonic crystals (PCs), consisting of a periodically patterned material, hold the potential for forming high-$Q$ optical microcavities with modal volumes approaching a cubic half-wavelength in the host material, the smallest theoretical volume.  Planar two-dimensional (2D) PC slab waveguides are a particularly attractive geometry due to the scalability of the planar fabrication process used, as well as the flexibility in the tailoring of cavity mode anti-node positions, polarization, radiation direction, and wavelength through adjustments in the lattice geometry\cite{ref:Painter1}.  Since the first demonstration of lasing in the defect of a planar 2D PC slab waveguide\cite{ref:Painter3}, much effort in design and fabrication \cite{ref:Vuckovic2,ref:Srinivasan1,ref:Ryu3,ref:Loncar3,ref:Yoshie2} has been extended to improve the $Q$-factor of planar PC microcavities.  In order to take advantage of the ultra-small mode volume and design flexibility afforded by planar PC cavities in applications such as cavity QED\cite{ref:Khitrova,ref:Vernooy2}, single photon on demand sources\cite{ref:Michler,ref:Santori}, or non-linear optical elements\cite{ref:Spillane1}, and to make them competitive with other small mode volume microcavities such as microdisks\cite{ref:Gayral} and microposts\cite{ref:Gerard2}, a $Q \gtrsim 10^4$ is required.  Although a number of designs have been proposed with predicted $Q$s in excess of $10^4$ \cite{ref:Vuckovic2,ref:Ryu3}, experimentally demonstrated values have been limited to around $2,800$ \cite{ref:Yoshie2}, with an increase to $4,000$ for larger mode volume designs where the defect consists of several missing air holes \cite{ref:Reese}.       

\begin{figure}[t]
\begin{center}
\epsfig{figure=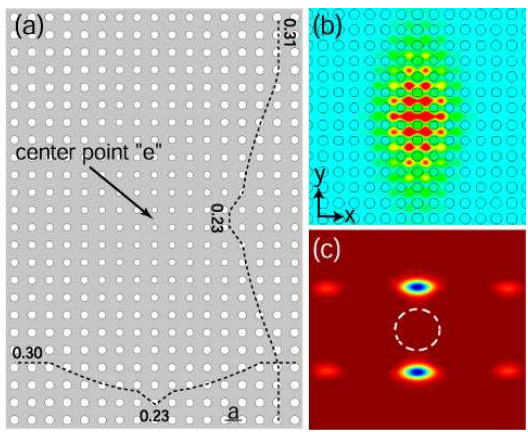, width=0.8\linewidth}
\caption{(a) Graded square lattice designed in \cite{ref:Srinivasan1}; dotted lines show the grade in hole radius ($r/a$) along the central $\hat{x}$ and $\hat{y}$ axes of the cavity. For the $A_{2}$ mode ($a/\lambda_o=0.25$): (b) Magnetic field amplitude ($B_z$) in the center of the PC membrane, and (c) Fourier transformed dominant electric field component ($E_{x}$). The dashed circle in (c) denotes the cladding light cone, showing that vertical radiation has been significantly suppressed.}
\label{fig:sq_graded_lattice}
\end{center}
\end{figure}

In Ref. \cite{ref:Srinivasan1}, high-$Q$ PC cavities were designed by considering the Fourier space properties and symmetry of cavity modes.  An important feature of these designs is their robustness, in that perturbations to the size and shape of \emph{individual} holes do not deteriorate the $Q$ significantly.  Vertical radiation losses, which are characterized by the presence of power at in-plane momentum components ($\bf{k_{\perp}}$) that lie within the light cone of the slab waveguide cladding, are reduced by choosing modes of a specific symmetry.  In particular, the modes selected are those that are odd about mirror planes normal to the direction of the mode's dominant Fourier components, eliminating the DC ($\bf{k_{\perp}}=0$) part of the field.  For the square lattice cavities studied in Ref. \cite{ref:Srinivasan1}, one such mode is a donor-type mode (labelled $A_{2}$ due to its symmetry) centered in the dielectric between two rows of air holes (point $e$ in Figure \ref{fig:sq_graded_lattice}).  Further improvements to both the in-plane and vertical loss are achieved by grading the lattice as shown in Figure \ref{fig:sq_graded_lattice}(a).  Figure \ref{fig:sq_graded_lattice}(b)-(c) shows the magnetic field amplitude and Fourier transformed dominant electric field component for the resulting $A_{2}$ mode as calculated by finite-difference time-domain (FDTD) simulations.  FDTD calculations predict $Q \sim 10^5$ for this mode, with a modal volume $V_{\text{eff}} \sim 0.25$ cubic half-wavelengths in air ($9.8$ $(\lambda/2n)^3$ for refractive index $n=3.4$, $\sim 4\text{X}$ larger than original designs in Ref. \cite{ref:Painter3}).  Calculations show that the grade used in Figure \ref{fig:sq_graded_lattice}(a) can be varied fairly significantly without degrading the $Q$ to a value less than $\sim 2{\text{x}}10^4$.

\begin{figure}[t]
\begin{center}
\epsfig{figure=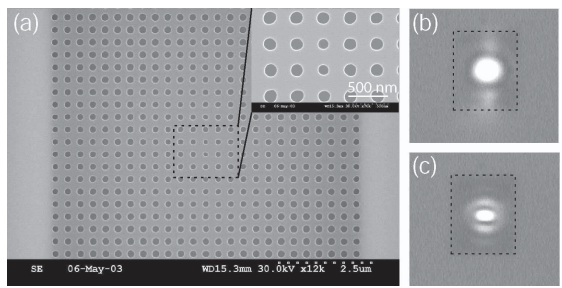, width=0.85\linewidth}
\caption{(a) SEM image of graded lattice PC microcavity in InAsP multi-quantum well material.  Lattice constant $a\sim305$ nm, membrane thickness $d=252$ nm. Optical image of cavity pumped with a (b) diffuse beam and (c) focused beam (dashed rectangle represents the physical extents of the undercut PC).}
\label{fig:fab_SEM}
\end{center}
\end{figure}

In order to measure the properties of the donor-type $A_{2}$ mode, graded square lattice PC cavities were fabricated in an active material consisting of five InAsP compressively-strained quantum wells, with peak spontaneous emission at 1285 nm.  The details of the epitaxial growth and some of the important properties of the materials system are reported in Ref. \cite{ref:Hwang2}.  The creation of the 2D PC membrane is accomplished through a number of steps, including electron-beam lithography, pattern transfer to a SiO$_2$ mask using an inductively coupled plasma reactive ion etch (ICP/RIE), and a high-temperature ($205$ $^{\text{o}}$C) Ar-Cl$_2$ ICP/RIE etch through the active material into a sacrificial InP layer.   The sample is undercut by removing the InP layer with a HCl:H$_2$O (4:1) solution leaving a 252 nm thick free-standing membrane; scanning electron micrographs (SEMs) of the graded lattice cavity are shown in Figure \ref{fig:fab_SEM}.  Each cavity consisted of a total of 32 rows and 25 columns of air holes, with a lattice spacing of $a=305$, $315$, $325$, or $335$ nm (chosen for $1.3$ $\mu$m emission to be close to the normalized frequency of the $A_{2}$ mode, $a/\lambda_o\sim0.25$), for total cavity dimensions on the order of $8\text{x}11 \mu$m.  The designed grade produces holes with radii between $r=70$-$110$ nm.  As will be described elsewhere, the fabrication was optimized to produce smooth vertical sidewalls in the holes, and to obtain small air holes, both of which were limitations of previous PC cavity work\cite{ref:Painter3}.       

\begin{figure}[ht]
\begin{center}
\epsfig{figure=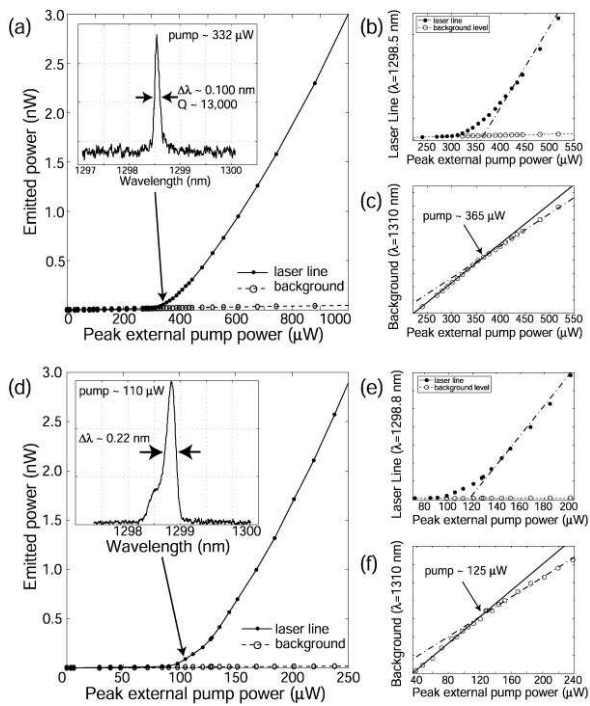, width=0.9\linewidth}
\caption{(a) L-L curve and sub-threshold spectrum (inset) of a graded square lattice PC microcavity pumped with a spatially broad pump beam (10 ns pulse, 300 ns period), and zoomed-in plots of (b) laser threshold and (c) background emission for the same diffuse pump beam.  (d) L-L curve and sub-threshold spectrum (inset) taken with a focused pump beam spot, showing increased thermal broadening in comparison to the diffuse pump beam spectrum, and zoomed-in plots of (e) laser threshold and (f) background emission.  The ``guide'' lines displayed in (b), (c), (e), and (f) are least-squares fits of the data taken over several points above and below the lasing transition region.}
\label{fig:diffuse_focused_pump_L_L}
\end{center}
\end{figure}

Devices are optically pumped (typically with a $10$ ns pulse width and $300$ ns period) at room temperature with a semiconductor laser at $830$ nm through a 20X objective lens, also used to collect emitted photoluminescence (PL) into an optical spectrum analyzer (OSA).  We initially pump the cavities with a broad pump beam (see Figure \ref{fig:fab_SEM}(b), area $\sim 21 \mu\text{m}^2$) for two reasons: (i) the broad pump beam covers a significant portion of the cavity area, so that after diffusion of carriers, the majority of the cavity should be pumped and therefore non-absorbing, and (ii) use of a broad pump beam limits the effects of thermal broadening, which, as discussed below, are significant for focused pump beams.   A typical L-L (light-in vs. light-out) curve using the broad pump beam condition is shown in Figure \ref{fig:diffuse_focused_pump_L_L}(a), where the power in the laser line is taken over a 10 nm bandwidth about the laser wavelength of $\lambda=1298.5$ nm.  In addition, the off-resonance background emission at $\lambda=1310$ nm was measured over a similar 10 nm bandwidth.  For low pump powers ($<300$ $\mu$W), the off-resonance emission and resonant wavelength emission linearly increase with pump power and are essentially identical in level, i.e., no resonance feature is observed.  Above  300 $\mu$W, we just begin to see a resonance peak in the spectrum and a characteristic super-linear transition from below threshold to above threshold follows.  In order to estimate the position of threshold we extrapolate back the L-L curve from above threshold (Fig. \ref{fig:diffuse_focused_pump_L_L}(b)), giving an approximate threshold pump level of $360$ $\mu$W.  A plot of the off-resonance emission (Fig. \ref{fig:diffuse_focused_pump_L_L}(c)) shows a (weak) slope change around $365$ $\mu$W giving a similar value for the estimated threshold value.  The slope change in the laser line versus pump power is initially a result of the material becoming transparent (more photons being radiated as opposed to being absorbed), and is then due to an increase in the stimulated rate of emission into the cavity mode as the lasing threshold is crossed, increasing the radiative efficiency into the cavity mode due to the presence of non-radiative carrier recombination and spontaneous emission into other modes.  The kink in the off-resonance background emission L-L curve can be attributed to the clamping of the carrier density (gain) in the region of the cavity mode and consequent saturation of the off-resonance (non-lasing modes') emission.  The background emission continues to increase after crossing threshold (rather than completely saturating) as a result of the pumping of areas which are outside of the cavity mode volume and thus not affected by the gain clamping (non-equilibrium carrier distributions\cite{ref:Slusher1} may also play a role). 

In order to estimate the cold cavity $Q$ value of the PC microcavity mode we measured the linewidth of the resonance in the PL around threshold.  The full-width half-maximum (FWHM) linewidth narrows from $0.138$ nm (at the lowest pump level we could accurately measure the linewidth, $320$ $\mu$W) down to $0.097$ nm at threshold.  A simple steady state rate equation model\cite{ref:Agrawal} of the cavity photon and excited state populations estimates the threshold pump level (with this beam size) to be $\sim 350$ $\mu$W for $Q\sim10^4$ in this quantum well active material, close to the experimentally measured value.  In this model the transparency carrier density occurs within $10\%$ of the threshold carrier density for cavity modes with $Q>10^4$.  A PL spectrum (Fig. \ref{fig:diffuse_focused_pump_L_L}(a), inset) for this device with the broad pump conditions, measured soon after detection of a resonance feature in the spectrum and below the estimated threshold level by about $10\%$, shows a resonance linewidth $\Delta\lambda=0.100$ nm, corresponding to a best estimate of the cold cavity $Q\sim1.3\text{x}10^4$. Above threshold we do not see further linewidth narrowing due to the resolution limit of the scanning monochromator ($0.08$ nm) used, as well as the presence of thermal broadening of the emission line during the pump pulse (this may also be partially due to incomplete saturation of the carrier density\cite{ref:Slusher1}).

By using a more tightly focused beam (see Figure \ref{fig:fab_SEM}(c), area $\sim8$ $\mu$m$^2$), the lasing threshold is considerably reduced.  In Figures \ref{fig:diffuse_focused_pump_L_L}(d)-(f), we plot the L-L curve for the laser line and off-resonance background emission using such a pump beam.  The plots are qualitatively similar to those for the diffuse pump beam; we begin to see a resonance feature when the pump power exceeds 95 $\mu$W.  Estimates for the threshold pump power from the laser line curve and off-resonance background emission are 120 $\mu$W and 125 $\mu$W, respectively.  The reduction in lasing threshold from the broad pump beam to the focused pump beam follows a nearly linear scale with the area of the pump beam.  Through further optimization of the pump beam, lasers with thresholds as low as $\sim 100$ $\mu$W have been observed.  From the sub-threshold spectrum shown in the inset of Figure \ref{fig:diffuse_focused_pump_L_L}(d) it is readily apparent that the lineshape has thermally broadened (the measured linewidth is now $0.220$ nm), as evidenced by its asymmetric shape on the short wavelength side.  To reduce the effects of this thermal broadening, the duty cycle can be decreased to $1\%$ (1 $\mu$s period and 10 ns pulse width), resulting in a less asymmetric resonance and sub-threshold linewidth of approximately $0.13$ nm.  Conversely, we have also increased the duty cycle to $25\%$ (1 $\mu$s period and 250 ns pulse width) and still observe lasing; heating in the membrane precludes lasing at even higher duty cycles.  

\begin{figure}[t]
\begin{center}
\epsfig{figure=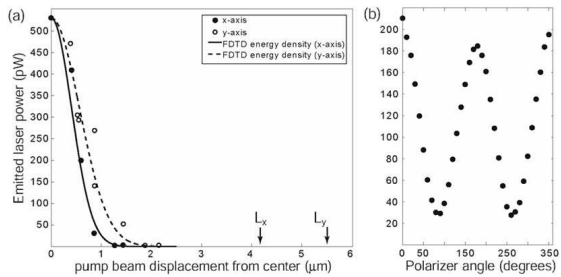, width=0.85\linewidth}
\caption{(a)  Emitted laser power as a function of pump position along the $\hat{x}$ and $\hat{y}$ axes of the cavity.  FDTD-generated Gaussian fits to the \emph{envelope} of the electric field energy density of the cavity mode are shown for comparison (note that the effective mode volume is calculated from the \emph{peak} electric field energy density).  L$_{x}$ and L$_{y}$ correspond to the physical extent of the PC in the $\hat{x}$- and $\hat{y}$-direction, respectively.  (b) Emitted laser power as a function of polarizer angle with respect to the $\hat{x}$ axis of the cavity.}
\label{fig:mode_volume_pol}
\end{center}
\end{figure}

To determine whether the laser mode described above is indeed the localized $A_{2}$ mode of Figure \ref{fig:sq_graded_lattice}, we have measured polarized intensity in the far-field as well as the sensitivity of the emitted laser power to pump position.  The measurements show the mode to be predominantly polarized along the $\hat{x}$-axis (Fig. \ref{fig:mode_volume_pol}(b)) of the cavity, consistent with FDTD results, and eliminating the possibility that the mode is of the other potential symmetry supported by the cavity \cite{ref:Srinivasan1}.  Furthermore, the lasing mode discussed above is the longest wavelength mode observed in the devices tested (higher frequency resonances are observed in some detuned devices), suggesting that it is the fundamental mode shown in Figure \ref{fig:sq_graded_lattice}(b), and not a higher order version of it.  In Figure \ref{fig:mode_volume_pol}(a) we show measurements of the emitted laser power as a function of the pump beam position (taken to be the center of the beam) relative to the center of the cavity (uncertainty in the pump position is $\sim 0.25$ $\mu$m).  The measurements indicate the mode is highly localized within the center of the cavity, consistent with simulations.

In summary, we have observed linewidths of $\Delta\lambda=0.10$ nm, corresponding to a cavity $Q$ of $1.3\text{x}10^4$, in sub-threshold measurements of graded square lattice photonic crystal microcavity lasers fabricated in an InAsP/InGaAsP multi-quantum well membrane.  In addition, lasing is seen at threshold peak external pump powers as low as 100 $\mu$W.  Measurements of the emitted power as a function of pump position show the mode to be strongly localized and give an estimate of the modal localization that is consistent with FDTD results.  This realization of a high $Q$, small mode volume microcavity is an important step in demonstrating the potential of PC microcavities for use in optoelectronics and quantum optics.     

K. Srinivasan thanks the Hertz Foundation for its financial support.  

\bibliography{/home/kartik/PBG_bibliography/PBG}
\end{document}